\documentclass[preprints,article,accept,moreauthors,pdftex,10pt,a4paper]{Definitions/mdpi}

\usepackage{amsmath,amssymb}
\usepackage[normalem]{ulem}

\firstpage{1} 
\pubvolume{xx}
\issuenum{1}
\articlenumber{5}
\pubyear{2018}
\copyrightyear{2018}
\history{Received: 30 November 2019; Accepted: 13 December 2019; Published: 16 December 2019}

\Title{Thermal and Quantum Fluctuation
~Effects in Quasiperiodic Systems in External Potentials}

\Author{Fabio Cinti $^{1,2}$ and Tommaso Macr\`i $^{3}$}

\address{$^{1}$ \quad Department of Physics and Astronomy, University of Florence, Via Sansone 1, I-50019, Sesto Fiorentino (FI), Italy; fabio.cinti@unifi.it

$^{2}$  \quad Department of Physics, University of Johannesburg, P.O. Box 524, Auckland Park 2006, South Africa.

$^{3}$ \quad Departamento de F\'isica Te\'orica e Experimental, International Institute of Physics, Universidade Federal do Rio Grande do Norte, Lagoa Nova, RN, 59078-970, Brazil; macri@fisica.ufrn.br}

\abstract{We analyze the many-body phases of an ensemble of particles interacting via a Lifshitz--Petrich--Gaussian pair potential in a harmonic confinement. We focus on specific parameter regimes where we expect decagonal quasiperiodic cluster arrangements.  Performing classical Monte Carlo as well as  path integral quantum Monte Carlo methods, we numerically simulate systems of a few thousand particles including thermal and quantum fluctuations. Our findings indicate that the competition between the intrinsic length scale of the harmonic oscillator and the wavelengths associated to the minima of the pair potential generically lead to a destruction of the quasicrystalline pattern. Extensions of this work are also discussed.}


\begin{document}

\section{Introduction}
Quasicrystals are quasiperiodic systems which break translational symmetry 
and display long-range order without being periodic \cite{PhysRevLett.53.1951}. Crystallographic theorem
in both two and three dimensions allows only a limited number of discrete symmetries
in the Bragg spectrum, namely two- three-, four-, and six-fold symmetries \cite{2002}. 
Other discrete symmetries for periodic systems, such as the well known five-fold (pentagonal) 
and ten-fold (decagonal), are prohibited.
A microscopic mechanism to generate a host of different
phases including quasicrystals is based on
simultaneous instability at more than one length scale, corresponding to
degenerate minima of the Fourier transform of the potential \cite{Likos01b,Shin09}. 

Typically, the instability at a single length scale for an isotropic potential 
is an indication of crystallization toward a regular lattice (Wigner crystal), which in two dimensions
leads to a triangular arrangement of particles. This is indeed the case of generic
power-law interactions, such as the van der Walls potential \cite{Fetter}.

It is important to note that power-law interactions are singular at the origin, 
prohibiting the approach of two or more particles at very small interparticle distances,
as in the case of dipolar potentials \cite{Lahaye2009,PhysRevA.93.061603}.
To circumvent this effect, one may engineer different types of interactions which are 
finite for small separations. This is the case of soft-core interactions that can be encountered
in a variety of setups, from Rydberg-dressed atoms \cite{Henkel2012,PhysRevA.87.061602,Cinti:2014aa,PhysRevA.89.011402,Macri:2014aa} to soft-matter systems \cite{Butenko2009,Likos2001,PhysRevLett.118.067001,PhysRevE.92.052307}.
At the many-body level, a finite interaction at small distances allows the formation
of cluster (or droplet) crystalline phases \cite{PhysRevLett.119.215302,PhysRevA.96.013627,Cinti2019,Macr2019,PhysRevLett.123.015301,1367-2630-16-3-033038,PhysRevLett.108.175301,PhysRevA.97.053623,PhysRevA.98.023618,PhysRevA.95.023622}.  
Each cluster generically contains a few  particles to several tens of particles. 
At the quantum level, such arrangements can lead 
to genuinely novel effects. A significant instance is provided by the supersolid phase, which
spontaneously breaks both translational symmetry and global gauge symmetry.
Supersolid phases have recently been  realized
in dipolar systems \cite{Kadau2016,PhysRevX.6.041039,Tanzi2019}, atom--light coupled cavity experiments \cite{Lonard2017} and spin-orbit
coupled BECs
~\cite{Li2017}.

In this work we study the many-body phases for an ensemble of distinguishable particles interacting 
via a Lifshitz--Petrich--Gaussian pair potential  in the presence of an external trapping potential. Extensive analysis of the phases has been 
done in the classical regime \cite{Barkan14,Dotera2014}, and more recently also in the quantum limit \cite{1905.12073}. Examples
of phases for these potentials include various types of quasicrystalline order and stripe phases.

Here we focus on a restricted parameter regime where we expect a 
decagonal quasiperiodic arrangement of clusters in free space, and then study the effect on
the many-body phases, adding a trapping potential. 
To explore such phases, we employ either classical Monte Carlo simulations to analyze the effect of 
thermal fluctuations or path integral quantum Monte Carlo methods for the investigation of
quantum fluctuations. Interestingly, we find that the combination of an external trapping potential and the thermal and quantum fluctuations destroy the decagonal cluster quasicrystal.

The rest of the paper is organized as follows: In Section \ref{section2} we introduce the 
microscopic model, and give some details on the methodologies that were applied. 
Section \ref{section3}  outlines findings regarding the trapped system in a classical regime. 
In Section \ref{section4}, we illustrate the results related to the quantum case.
Our conclusions and extensions of this work are addressed in Section \ref{section5}.
 
 \section{Model Hamiltonians and Methodology}\label{section2}
In this section we introduce our system for both the classical and the quantum case. 
Considering a set of $N$ distinguishable particles of mass $m$ and trapped in a harmonic potential, 
the Hamiltonian reads: 
\begin{equation}
\label{Ham1}
H=-\lambda\sum_{i=1}^{N}\nabla^2_i+\sum_{i<j}^N V\left({\bf r}_{ij}\right) + \gamma\sum_	{i=1}^{N}{\bf r}_i^2\,,
\end{equation}
where the kinetic contribution to the total energy is scaled by the parameter $\lambda=\frac{\hbar^2}{2m}$, 
${\bf r}_i \equiv (x_i ,y_i )$ being the position of $i$-th particle on the plane, and ${\bf r}_{ij}  = | {\bf r}_i - {\bf r}_j |$.  $\lambda=0$ reflects a pure classical regime, whereas  $\lambda>0$ takes into account  
quantum fluctuations. The second sum represents the Lifshitz--Petrich--Gaussian pair potential \cite{Barkan14}, which can be written as:
\begin{equation}\label{eq:pot}
V(r) = e^{-\frac{1}{2}\sigma^2r^2}\left( c_0 + c_2 r^2 + c_4 r^4+ c_6r^6 + c_8  r^8\right)\,.
\end{equation}

\begin{figure}[H]
\centering
\includegraphics[width=15cm]{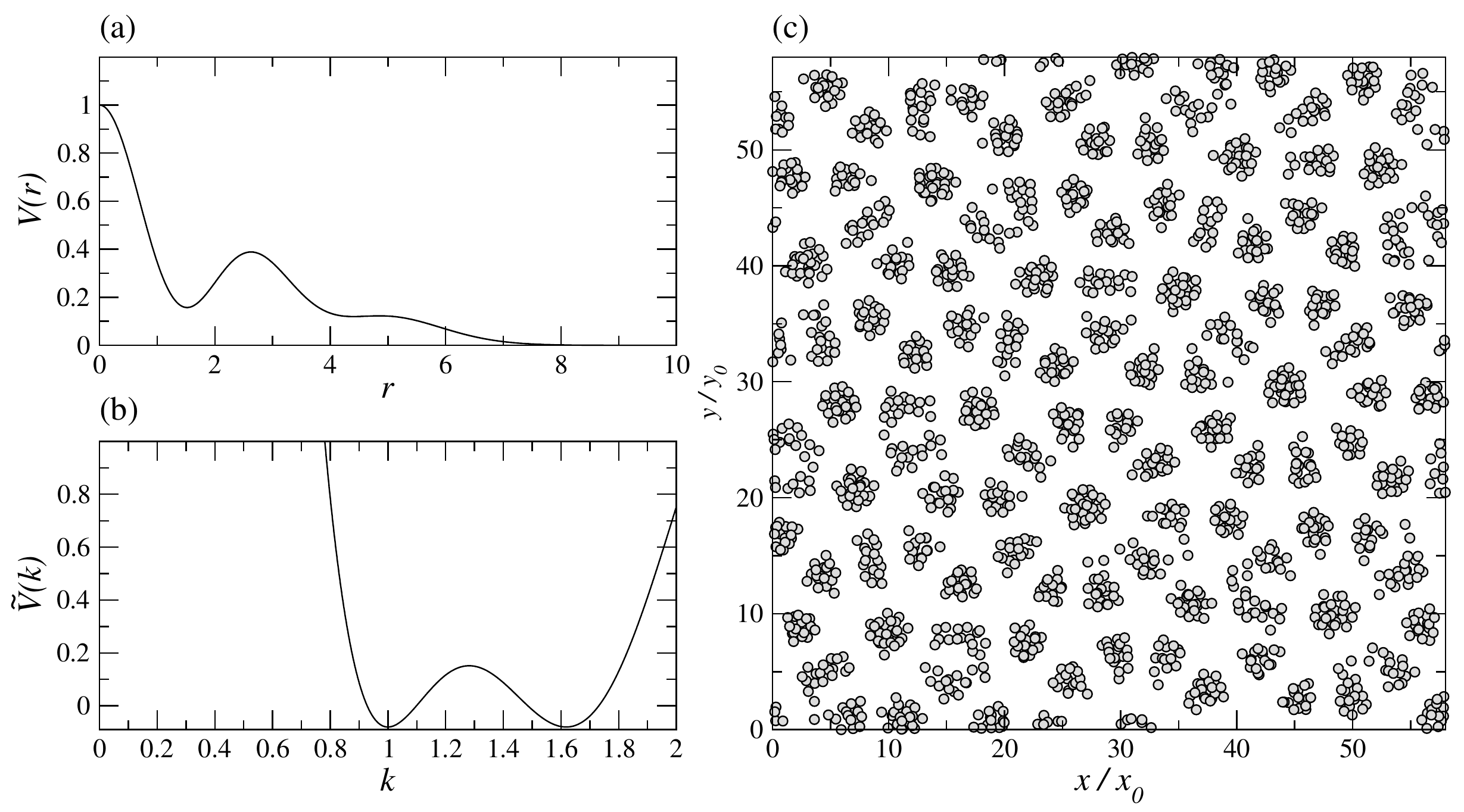}
\caption{(\textbf{a}) Pair potential in real space identifying a cluster quasicrystal with a 10-fold symmetry. 
(\textbf{b}) Fourier of the pair potential proposed in (a) around the minima that mark the quasicrystal.
(\textbf{c}) Typical configuration of a classical simulation obtained at $t=0.05$ using 2048 particles (canonical ensemble).
The parameters of the interparticle potential in Equation (\ref{eq:pot}) used for the simulations 
were $\sigma = 0.69$, $c_0 = 1$, $c_2 = -0.79$, $c_4 =  0.25$, 
$c_6 = -0.02$, and $c_8 =  6.0\times 10^{-4}$, as reported in Ref.~\cite{Barkan14}. These  parameters lead to two degenerate minima of the 
Fourier transform of the potential. Length scales and parameters were chosen to fix the first minimum at
$k_{min}^{1}=1$.  
}\label{figure1}
\end{figure}

The parameters $\sigma$ and $c_{i}$ have to be chosen such that the equilibrium configuration of the system without confinement establishes a quasicrystal structure below the transition temperature. 
The caption of Figure~\ref{figure1} reports the exact value of $\sigma$ and $c_{i}$ for such a quasicrystal pattern. The two-body potential furnishes two equal-depth negative minima in its Fourier transform. 
Since the present work aims to consider cluster crystals featuring a 10-fold symmetry,
the ratio of the corresponding wave-vectors yields $\left(1+\sqrt{5}\right)/2\approx1.618$ \cite{Barkan14}.  
Figure~\ref{figure1}a shows the pair potential, and its Fourier transform around the two equal-depth negative minima region is reported in Figure~\ref{figure1}b. The harmonic confinement of  Equation~\eqref{Ham1} is represented by the third term in Equation~(\ref{Ham1}), where $\gamma$ is the strength of the trap.

We studied the equilibrium state of Equation~(\ref{Ham1}) by varying the reduced temperature $t=k_BT/V_0$, $V_0=V(0)$, and $N$. To characterize the limit for $\lambda=0$ we employed standard classical Monte Carlo simulations. The numerics were initialized by choosing a random arrangement of particles. Thermodynamic equilibrium was first reached at a high temperature, $t_0$. Then the temperature was gradually decreased $t\to t - \delta t$ ($\delta t > 0$) starting with the last configuration sampled at the previous higher temperature. The procedure was completed when the final temperature was reached. 

First we neglected the trapped potential ($\gamma=0$) and used periodic boundary conditions along all directions in order to reproduce a quasicrystal cluster with a decagonal symmetry. Following Ref.~\cite{Barkan14}, we fixed the reduced density at $\rho r_0^2=0.8$, where $r_0=2\pi/k_\text{min}$ is the characteristic length given by the inverse of the wave vector corresponding to the first minimum, $k_\text{min}$, of the Fourier transform of Equation~\eqref{eq:pot}. As a result of the annealing process,  Figure~\ref{figure1}c shows a pattern characterized by a 10-fold symmetry at  temperature $t=0.05$ and $N=2048$.

Regarding the quantum counterpart ($\lambda>0$), we investigated the equilibrium properties of the system described by the Hamiltonian \eqref{Ham1} employing first-principle computer simulations based on a continuous-space path integral Monte Carlo \cite{Ceperley1995,krauth2006statistical}. A thorough illustration of the methodology can be found in \cite{Boninsegni2005}. Since the potential in Figure~\ref{figure1}a is well behaved, 
the density operator can be approximated using a fourth-order expansion, as proposed by Chin in Ref.~\cite{Chin1997}. 

\section{Trapped Quasicrystal: Thermal Fluctuations}\label{section3}
\begin{figure}[H]
\centering
\includegraphics[width=15cm]{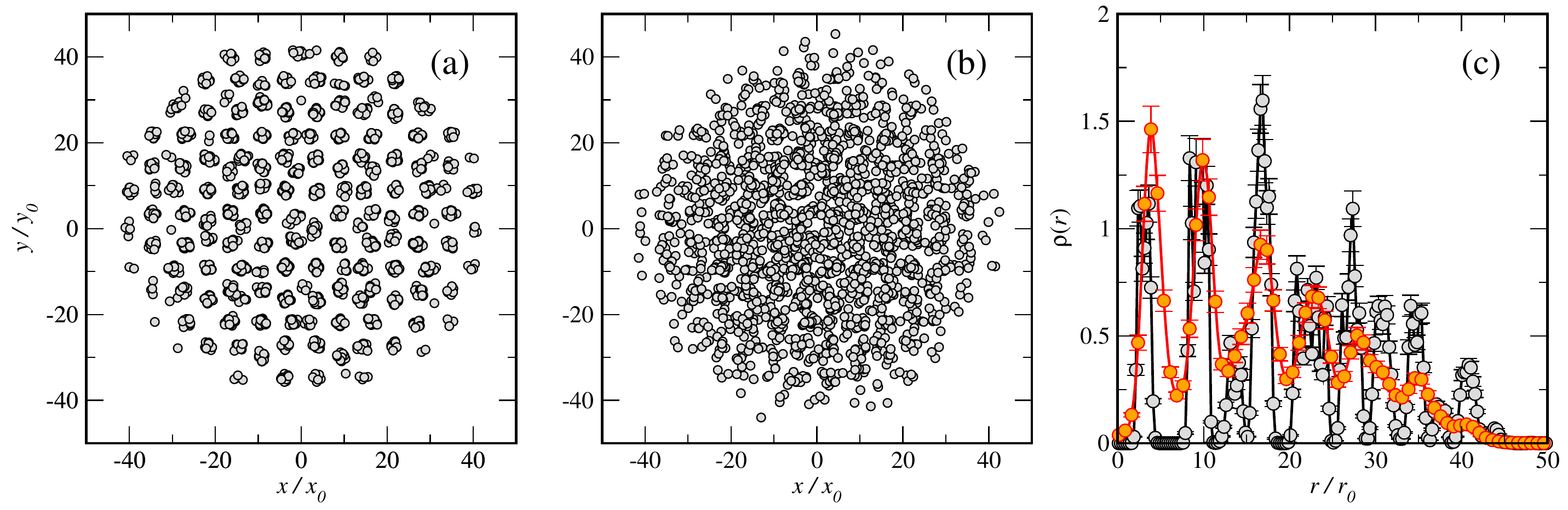}
\caption{Instantaneous configuration for a trapped system made of $N = 2048$ classical particles at temperature (\textbf{a}) $t=0.05$ and (\textbf{b}) $t=1.0$. (\textbf{c}) Grey points: density profiles for panel (a); Orange points: density profiles for panel (b).
}\label{figure2}
\end{figure}

Now we discuss the properties of the quasicrystal characterized by a $10$-fold symmetry in the presence 
of a two-dimensional harmonic confinement. As stated above, we started studying the system 
by excluding quantum effects, that is, imposing $\lambda=0$. Figure~\ref{figure2}a depicts a snapshot configuration of a classical 
Monte Carlo simulation run employing $N=2048$ and with strength $\gamma=0.01$ 
(in units of $V_0$) of the harmonic confinement. In Figure~\ref{figure2}a we  consider $t=0.05$.  
Each grey circle  represents the position of a particle on the $xy$-plane.
The wave-number, $k_c$, which denotes the harmonic trap, is introduced defining the characteristic length of the harmonic trap as  $l_c=\sqrt{\frac{V_0}{2\gamma}}\approx 7.07$, then it yields $k_c=\frac{2\pi}{l_c}\approx0.89$.

The structure in Figure~\ref{figure2}a displays a strong modification of the quasicrystal pattern, and it appears as the result of the strong frustration imposed by the harmonic confinement. Qualitatively, we might split the trapped quasicrystal in two regions. The first region within the radius $r/r_0\lesssim20$ is an amorphous structure that recalls the ring-shaped symmetry of Figure~\ref{figure1}c. 
In contrast to this \textit{inner shell}, clusters rearranged themselves
forming two visible circles at the trap's border $20\lesssim r/r_0\lesssim45$. 
Indeed, these observations bring us to the conclusion that the system is no longer a quasicrystal. We also notice that each cluster had about half the number of particles with respect to the ones in Figure~\ref{figure1}c. Furthermore, particles clumped in a tighter arrangement. Figure~\ref{figure2}b shows a configuration for the same system but at higher temperature $t=1$. 
At first glance, the simulation proposed in Figure~\ref{figure1}c simply spots the increase of the thermal fluctuations whose effect is to move the systems towards a fluid phase. Nevertheless, a detailed inspection shows that  the inner shell still displayed rings at $t=1$. More precisely, Figure~\ref{figure2}b reflects a thermal state where fluctuations made rings broader and spatially linked among themselves. In contrast, the outer shell seemed to be completely disordered.  

The radial density profile $\rho(r)$ of the configurations in Figure~\ref{figure2}a and b is represented in Figure~\ref{figure2}c. The grey circles represent $\rho(r)$ of panel a, whereas the orange circles refer to panel b. The radial density distribution of the classical system shows pronounced and narrow maxima with every other well-defined minima located at zero. This indicates that clusters were not thermally linked as described above.  
Interestingly, the radial density function at higher temperature (orange points) looks like  the position of the peaks were mainly left intact for $r/r_0\lesssim20$. In a different way, the oscillations of $\rho(r)$ have minima that do not touch the $x$-axis, leading to a structure where particles have a non-zero probability of percolating from cluster to cluster.

\section{Trapped Quasicrystal: Quantum Fluctuations}\label{section4}
\begin{figure}[H]
\centering
\includegraphics[width=15cm]{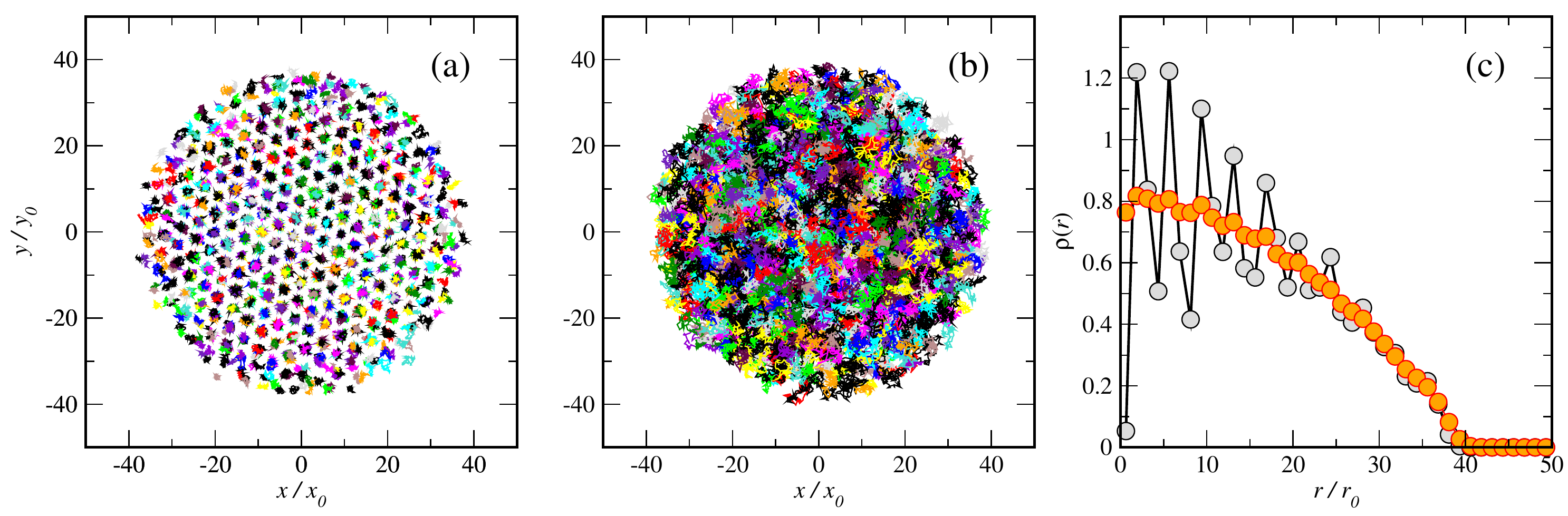}
\caption{Snapshot of $N=2048$ boltzmannons simulated at temperature $t=0.05$ and confined into an harmonic trap varying the kinetic term of \eqref{Ham1}: (\textbf{a})  $\lambda=0.05$ and (\textbf{b}) $\lambda=0.5$. Different colors identify different world lines. (\textbf{c}) Grey points: density profiles for panel (a); Orange points: density profiles for panel (b)---error bars lie within point size.
}\label{figure3}
\end{figure}

In the previous section we focussed on the Hamiltonian \eqref{Ham1} without quantum fluctuations. In the present section we analyze the model of Equation~\eqref{Ham1} through a path integral Monte Carlo approach that includes quantum fluctuations. 
In particular, we fix $t=0.05$ anew and vary $\lambda$ of \eqref{Ham1}. 
Figure~\ref{figure3} shows results with $\lambda=0.05$ (panel a) and $\lambda=0.5$ (panel b), respectively. Again energy scales are expressed in units of $V_0$. 
The strength of the trap was fixed at $\gamma=0.01$ as in the classical case. 
The wave-number specifying the strength of the trap now reads $k_q=\frac{2\pi}{l_q}$, $l_q=\left(\frac{2\lambda}{\gamma}\right)^{1/4}$ being the quantum harmonic oscillator length.
We point out that the choice of scaling the parameters in terms of the strength of the interparticle potential
$V_0$, and therefore changing the strength of the kinetic energy term, is arbitrary and reflects a convenience for numerical simulations without affecting the physical model.

Our simulations were performed without introducing bosonic exchanges, taking only classical statistics into consideration. However, such an approximation still accounts for zero-point motion effects
which are due to quantum fluctuations. Panels a and b show snapshots (instantaneous configurations) of the projection of world lines onto the $xy$ plane obtained by integrating over the imaginary time evolution. Different colors refer to different particles (within a quasi-classical approximation those particles are sometimes named $boltzmannons$). For a discussion of the properties of our simulations in this regime, see  the review by D. Ceperley \cite{Ceperley1995}.
It is important to stress that these kinds of projections are usually considered as a good representation of the square of the semi-classical many-body wave function. Figure~\ref{figure3}a shows particles whose paths remain confined on a single cluster. It is clear that, excluding clusters confined on the edge of the trap, the configuration results to form a perfect triangular cluster crystal. The same information can be drawn by looking at the radial density profile in Figure~\ref{figure2}c. The grey points characterize density $\rho(r)$, which displays oscillations that are consistent with the arrangement of particles.

We can conclude that moderate small quantum fluctuations drive the systems toward a perfectly ordered solid consisting of evenly spaced multiple-occupancy clusters. In addition, the strong localization of paths seems to exclude the presence of a possible supersolid phase. Upon increasing the kinetic term (as in Figure~\ref{figure3}b), boltzmannons start to delocalize throughout the trap with a consequent quantum melting of the triangular crystal into a superfluid. The orange points in Figure~\ref{figure3}c account for this uniform superfluid state \cite{PhysRevB.84.014534}. 

\begin{figure}[H]
\centering
\includegraphics[width=10cm]{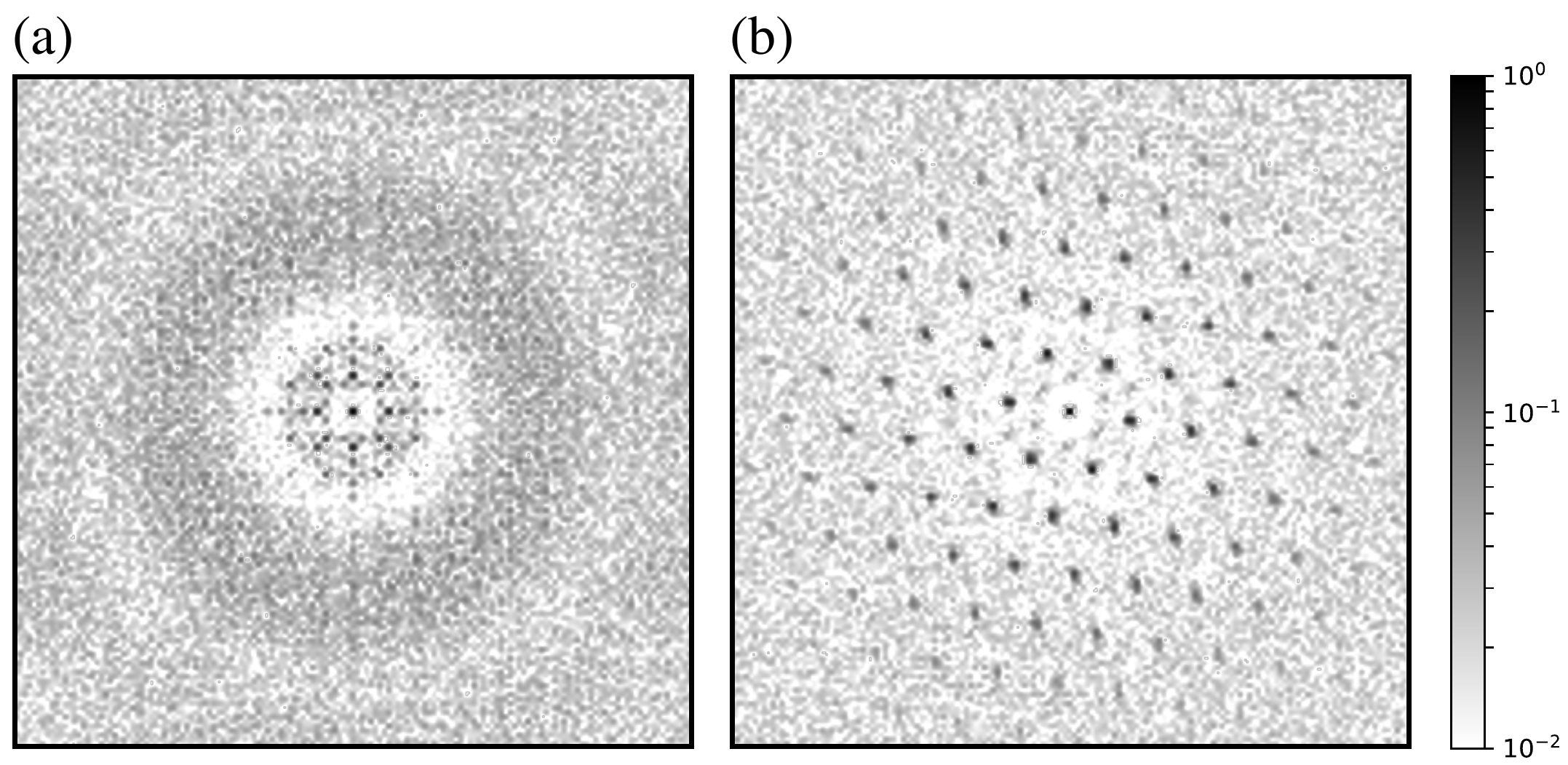}
\caption{Fourier transform (logarithmic scale) of the density for the classical and quantum configurations.
(\textbf{a}) Fourier transform of Figure \ref{figure2}a (thermal fluctuations), (\textbf{b}) Fourier transform of Figure \ref{figure3}a (quantum fluctuations). 
Whereas in (b) the peak structure clearly indicates the presence of a hexagonal lattice, the Fourier transform of the classical configuration in (a) displays a central peak surrounded by four pairs of closely spaced peaks, compatible with an irregular structure of the clusters.   
}\label{figure4}
\end{figure}   

Finally, we compare the structures obtained for the classical case in Figure~\ref{figure2}a and with the introduction of quantum fluctuations in Figure~\ref{figure3}a by looking at their Fourier transforms. Figure~\ref{figure4}a shows the two-dimensional Fourier transform of the classical configuration whereas Figure~\ref{figure4}b corresponds to the quantum case. The quantum case evidently confirms the onset of a hexagonal cluster crystal symmetry. Figure~\ref{figure4}b was calculated by considering position of the centroid coordinates. In contrast to the quantum case, the Fourier transform of the classical configuration displays a central peak surrounded by four pairs of closely spaced peaks.

\section{Conclusion}\label{section5}
In this work we studied the many-body phases of an ensemble of distinguishable particles in the presence of a Lifshitz--Petrich--Gaussian pair potential in a harmonic trap. 
The competition between the intrinsic length scale of the harmonic oscillator and the wavelengths 
associated to the minima of the interaction potential leads to a destruction of the quasicrystalline pattern.
We analyzed this effect both in the presence and  absence of quantum fluctuations for
particles obeying classical statistics.
We found that thermal fluctuations smeared out the fluctuations of the maxima and the minima 
of the radial density. On the other hand, weak quantum fluctuations induced a hexagonal density pattern, also
verified via the study of the Fourier transform of the density. 
Stronger fluctuations led to a transition to a fluid phase. 
The detailed investigation of particle exchanges for the characterization of 
superfluidity will be done in a separate study. Superfluid features are relevant to the identification of parameter regimes
where the simultaneous  breaking of translational and global gauge symmetry lead to an exotic 
supersolid phase \cite{RevModPhys.84.759,PhysRevLett.105.135301}. Additional work may also include the comprehensive analysis of finite size effects 
as a function of particle number and the trap strength, which may lead to different arrangements and occupations of the clusters, the study of collective excitations for single- and multi-component systems of bosons or fermions \cite{PhysRevA.81.033624}.

\vspace{6pt}
\authorcontributions{Both authors contributed equally. Both authors have read and agreed to the published version of the manuscript.}

\funding{ T.M. acknowledges the support of CAPES through the Project CAPES/Nuffic n.
88887.156521/2017-00 and CNPq through Bolsa de produtividade
em Pesquisa n. 311079/2015-6. This work was supported by the Serrapilheira Institute (Grant No. Serra-1812-27802 to T.M.)}

\conflictsofinterest {
The authors declare no conflict of interest. The founding sponsors had no role in the design of the study; in the collection, analyses, or interpretation of data; in the writing of the manuscript, or in the decision to publish the results.
} 

\reftitle{References}



\end{document}